\documentclass[acmsmall]{acmart}

\RequirePackage[normalem]{ulem}

\AtBeginDocument{%
  }

\begin{document}

\title[Understanding Phone-Carrying Behaviors of Wheelchair Users for Mobile Context-Awareness]{Not Just Pockets: Understanding Phone-Carrying Behaviors of Wheelchair Users for Mobile Context-Awareness}

\author{Yunzhi Li}
\affiliation{%
    \institution{Carnegie Mellon University}
    \city{Pittsburgh}
    \state{PA}
    \country{USA}
 }
\email{yunzhil@cs.cmu.edu}

\author{Patrick Carrington}
\affiliation{%
    \institution{Carnegie Mellon University}
    \city{Pittsburgh}
    \state{PA}
    \country{USA}
 }
\email{pcarrington@cmu.edu}

\begin{abstract}
Smartphone-based context-awareness holds significant promise for wheelchair users---from detecting everyday accessibility barriers to enabling ability-based adaptations. Such capabilities often build on passive context inference through mobile sensing, yet their accuracy hinges on how and where phones are carried and the resulting signal quality. While prior work documents phone-carrying behaviors in the general population, patterns specific to wheelchair users remain underexplored. Through a mixed-methods approach combining a survey of 91 and interviews with 15 wheelchair users, we systematically investigate their phone-carrying locations and influencing factors. Our findings reveal distinct patterns extending beyond pocket storage to diverse wheelchair-mounted accessories and around-body placements, shaped by the interplay of physical ability, wheelchair design, and everyday contexts, including social, activity, and device factors. Grounded in these findings, we articulate how carrying location can serve as a proxy for user context to enable novel context-aware experiences, and discuss design implications for developing inclusive and effective mobile context-aware applications.
\end{abstract}

\setcopyright{cc}
\setcctype{by}
\acmJournal{PACMHCI}
\acmYear{2026} \acmVolume{10} \acmNumber{5} \acmArticle{MHCI1226}
\acmMonth{8} \acmDOI{10.1145/3821694}


\begin{teaserfigure}
  \includegraphics[width=\textwidth]{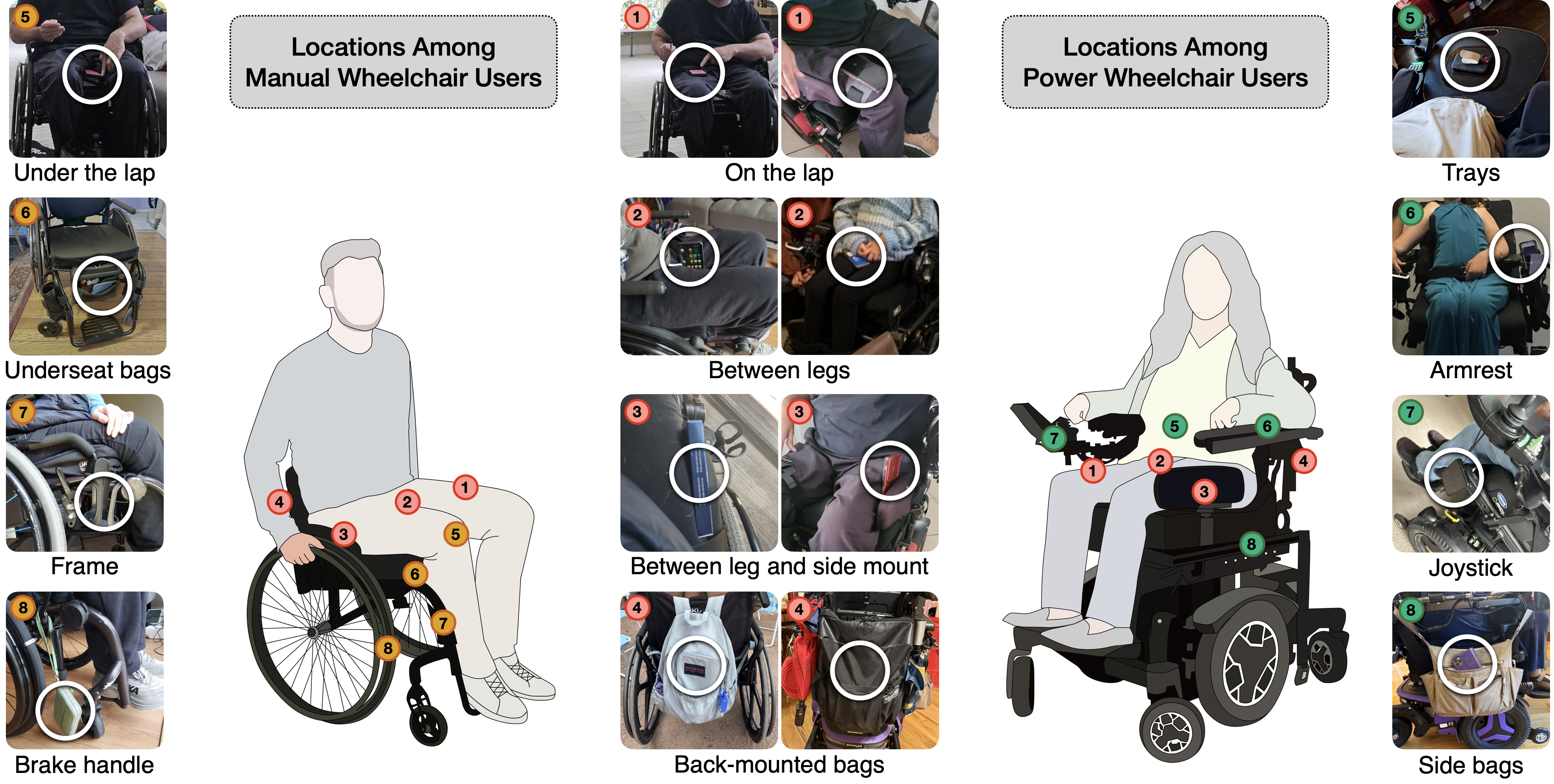}
  \caption{An illustration of diverse phone-carrying locations among wheelchair users, including common positions like on or between the legs and back-mounted bags, as well as wheelchair-specific placements, such as underseat, frame, or brake mounts on manual chairs, and armrest, joystick, or tray table mounts on power chairs. Colors are used to indicate location types, with red denoting common locations, orange denoting manual-specific, and green denoting power-specific.}
  \label{fig:teaser}
\end{teaserfigure}

\maketitle
\section{Introduction}
Context-awareness seeks to enable computing systems to perceive, interpret, and respond to their context of use for enhanced usability and user experience~\cite{dey2001understanding}.
Such capabilities are especially valuable for wheelchair users, as they allow systems to dynamically adapt to users' diverse abilities and mobility characteristics~\cite{wobbrock2018ability, wobbrock2019situationally} while offering personalized, just-in-time support.
Smartphones, with their widespread adoption and rich sensing capabilities, offer a practical and versatile platform for delivering such experiences, creating opportunities across numerous domains such as accessible navigation~\cite{gupta2020towards, wu2022alignment}, adaptive interaction techniques~\cite{mott2016smart, mott2019cluster}, inclusive activity recognition~\cite{ma2017activity, garcia2015identifying, hiremath2015detection}, and the development of smart wheelchairs~\cite{leaman2017comprehensive, zhang2022understanding, trivedi2013design}. 
Throughout this work, we use ``context'' to broadly refer to information concerning wheelchair users and their situations, encompassing both user factors (e.g., physical abilities, wheelchair type) and situational factors (e.g., activity, environmental, and social conditions).

A key to realizing these potentials lies in passively sensing user contexts through embedded sensors, particularly during non-interaction periods when devices are simply being carried~\cite{dey2001understanding, dey2018context}. 
However, the accuracy of context detection depends on the characteristics and quality of sensor signals, which vary significantly with phone placement~\cite{henpraserttae2011accurate, jackermeier2021smartphone, otim2019effects} and the user's mobility context~\cite{schaffer2017step, carroll2012use, carrington2015but}.  
For example, smartphone-based step counters and gait-monitoring applications are typically developed and validated using phones carried near the waist, such as in trouser pockets, during walking. When a phone is instead carried in a backpack or purse, these systems can silently fail to register meaningful activity or return erroneous parameters~\cite{werner2023validity, zhu2013apt}.
Similar placement sensitivities affect other context-aware capabilities, including fall detection~\cite{gjoreski2011accelerometer}, posture recognition~\cite{mollyn2023imuposer}, and sound classification~\cite{franke2009can}.
As a result, false assumptions about phone placement can systematically bias sensing and interaction, undermining system reliability and ultimately risking the exclusion of those who could benefit most~\cite{coskun2015phone, sztyler2016body, gao2019context, siean2021wearable}.

While prior research has documented phone storage patterns among the general population~\cite{wiese2013phoneprioception, redmayne2017s, dey2011getting, ichikawa2005s, zeleke2022mobile, sorysz2023beyond, nithiya2020mobile}, highlighting common locations such as pockets, purses, and bags, the phone-carrying behaviors and any associated influencing factors specific to wheelchair users remain largely unexplored.
Yet this gap merits closer examination, as wheelchair users tend to have distinct preferences, behaviors, and challenges with technology use when navigating physical and social environments~\cite{smith2016review, carrington2018exploring, jang2022should, sahoo2023wheelchair}.
Factors such as their unique mobility contexts, types of motor impairments, and use of assistive technologies all suggest that wheelchair users may adopt more diverse phone-carrying behaviors compared to the general population, potentially leading to increased complexity and additional technical challenges for context-aware applications. 

Recognizing this critical research gap, our study takes an initial step toward systematically examining phone-carrying practices among wheelchair users. To achieve both breadth and depth in our investigation, we adopt a mixed-methods approach, beginning with an online survey of 91 wheelchair users to uncover broad patterns and preferences, followed by semi-structured interviews with 15 participants to explore the factors influencing their choices in greater depth. Specifically, we focus on the following research questions throughout the study:

\begin{description}
    \item [RQ1:] What are the common phone-carrying locations among wheelchair users?
    \item [RQ2:] What factors influence wheelchair users' preferences for phone-carrying locations, and~why?
    \item [RQ3:] How can phone-carrying patterns among wheelchair users inform the design and development of context-aware mobile applications? 
\end{description}

Our findings reveal that wheelchair users adopt a wide range of phone-carrying practices that extend beyond conventional pocket and bag locations to include wheelchair-mounted accessories (e.g., holders, tray tables, underseat bags) and various around-body placements (e.g., between or under the lap). These practices are further shaped by considerations of reachability, safety, availability, social context, and the use of companion devices such as smartwatches. 
By uncovering these practices, our work addresses a fundamental yet largely underdocumented aspect of mobile interaction among wheelchair users and challenges the able-bodied assumptions that often underlie mobile computing research, where phones are typically presumed to be carried in pockets or held in hand.
We further show that phone-carrying behaviors among wheelchair users are not merely a matter of convenience but encode meaningful contextual information about their situational constraints, interaction intent, and mobility state. Building on these findings, we discuss implications for placement-aware interaction and cross-device coordination, context-aware physical accessory design, and inclusive development cycles that accommodate placement diversity to ensure application effectiveness. We also explore how authentic body- and chair-coupled carrying locations can create new opportunities for practical everyday sensing and context-awareness tailored to wheelchair users' needs.

To summarize, this paper makes the following contributions:
\begin{itemize}
    \item We provide the first systematic characterization of phone-carrying locations among wheelchair users through a mixed-methods study combining a survey ($N=91$) and interviews ($N=15$).
    \item We show how intrinsic and contextual factors shape these carrying practices, and synthesize these findings into a design framework that treats phone placement as a first-class design consideration for inclusive mobile context-awareness.
    \item We provide preliminary sensing demonstrations that illustrate both the impact of placement diversity on existing sensing assumptions and the feasibility of leveraging wheelchair-specific carrying locations for novel context-aware applications. Our sensing experiments are scoped as single-user proofs of concept to motivate future comprehensive validation.
\end{itemize}

\section{Background and Related Work}

In this section, we first provide background on wheelchair types and configurations to contextualize our findings. We then describe related work studying phone placement, general phone accessibility for people with motor impairments, and conclude with a discussion of the literature focusing on context-awareness for wheelchair users.

\subsection{Background: Wheelchair Types and Configurations}
Wheelchairs are broadly categorized into manual wheelchairs (MWs) and power wheelchairs (PWs). Manual wheelchairs are propelled by users through hand rims attached to the rear wheels or by attendants. The pushrim-propelled design is the most common configuration for everyday mobility and is used by the majority of manual wheelchair users. Other MW variants include sports-specific designs such as those used for basketball, tennis, or racing, and alternative propulsion systems such as crank- or lever-driven models, which are less common and typically used for exercise or specialized rehabilitation~\cite{flemmer2016review, sayed2024wheelchair}.
Power wheelchairs, in contrast, use electric motors typically located beneath the seat and are primarily controlled through a joystick interface. They vary widely in configurability, ranging from standard models to advanced systems featuring suspension, high-torque motors, or specialized seating options such as tilt-in-space or standing modes~\cite{hunt2004demographic}. PW designs also differ by drive wheel placement, including front-wheel, mid-wheel, and rear-wheel drives, each offering distinct handling characteristics suited to different environments~\cite{munakata2013five}.

Meanwhile, wheelchair accessories often fall into two categories: seating and positioning accessories and convenience accessories. Seating and positioning accessories, such as postural support components and safety equipment like pelvic positioning belts, are typically prescribed and installed during the clinical fitting process by healthcare professionals to support users' physical needs, safety, and mobility~\cite{cooper2006engineering}. In contrast, convenience accessories such as phone holders, cup holders, and tray tables are generally not included by default or addressed during clinical fitting. Instead, users select and add these items themselves based on their daily activities and personal preferences, often mounting them to components such as armrests, backrests, or side guards. This personalization reflects the highly individualized nature of wheelchair provision, where each setup is tailored to a user's physical abilities, daily activities, and environments, consequently shaping where and how smartphones can be carried or accessed.

\begin{figure}[h]
 \centering
\includegraphics[width=0.85\textwidth]{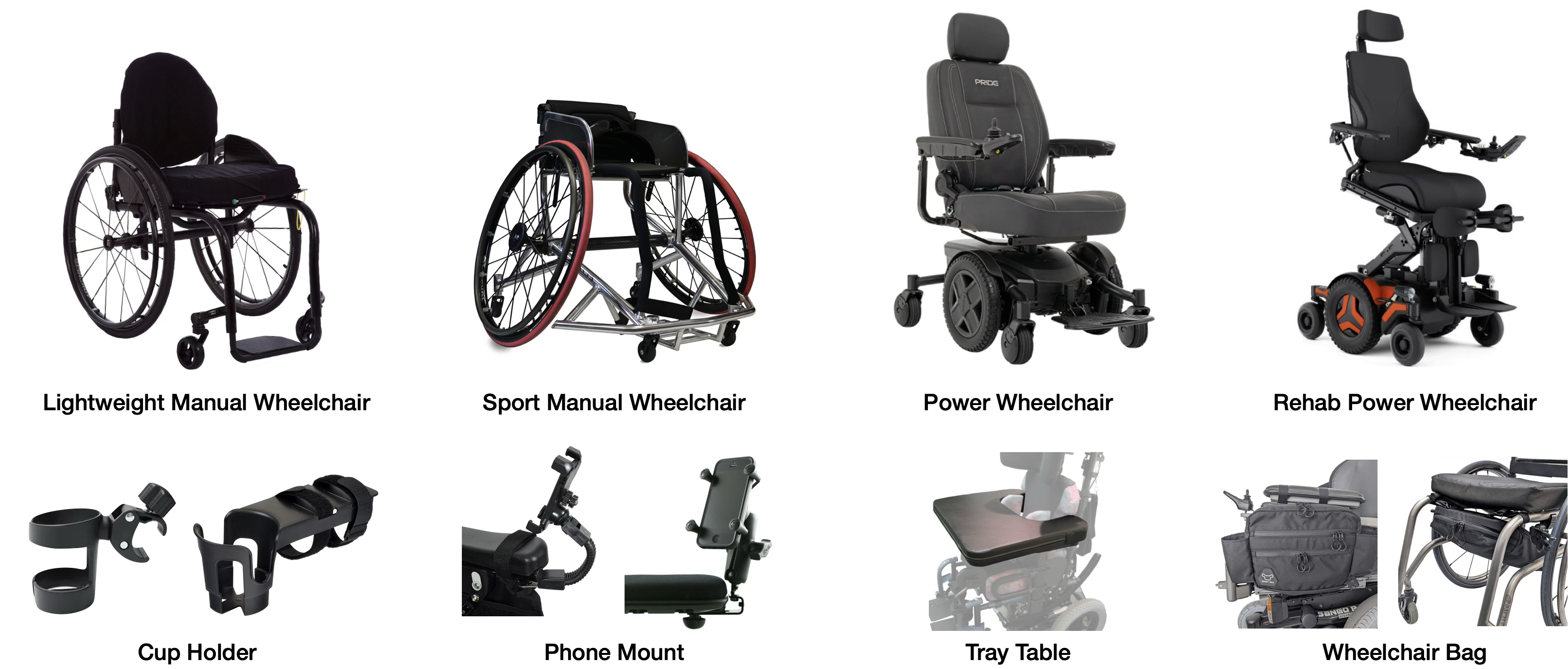}
\caption{Examples of common wheelchair types and accessories. The top row illustrates four wheelchair configurations: a lightweight manual wheelchair, a sport wheelchair with cambered wheels, a standard power wheelchair, and a complex rehab power wheelchair. The bottom row shows frequently used accessories, including cup holders, phone mounts, tray tables, and wheelchair bags.}
\label{fig:wheelchair_type}
\end{figure}

\subsection{Studies of Phone Placement and Their Implications on Context-Awareness}
Research on users’ behaviors around phone placement has been an active area of investigation, revealing insights into proximity, location preferences, and influencing factors. Early work by Patel et al.~\cite{patel2006farther} and Dey et al.~\cite{dey2011getting} examined how often individuals kept their phones within immediate proximity, finding that phones were within arm’s reach between 17\% and 85\% of the time. 
Beyond proximity, prior work has also studied the specific locations where people commonly carry their phones. Ichikawa et al.~\cite{ichikawa2005s} identified common locations such as trouser pockets, shoulder bags, and upper-body pockets, with preferences shaped by gender and age. Wiese et al.~\cite{wiese2013phoneprioception} extended this work through a mixed-methods study with 693 participants, identifying additional locations including purses and belt cases, and highlighting factors such as notification accessibility, safety concerns, and distraction minimization.

Such findings on phone placement patterns have enabled researchers to study the implications and influence of carrying location on context detection accuracy, revealing significant challenges for sensor-based systems. These investigations have shown that different carrying positions produce distinct accelerometer and gyroscope signatures for identical activities~\cite{lester2006practical, antos2014hand, stisen2015smart}, impacting applications such as step counting~\cite{zhu2013apt}, posture recognition~\cite{mollyn2023imuposer, gjoreski2011accelerometer}, activity detection~\cite{sun2010activity}, and energy expenditure estimation~\cite{manohar2011evaluation}. Other modalities are similarly affected: light sensors generate different readings when phones are in bags versus carried openly, introducing challenges for environmental recognition~\cite{yang2013efficient, yang2014sherlock}, while microphone data quality varies across positions, impacting ambient sound classification~\cite{franke2009can} and activity inference accuracy~\cite{lu2009soundsense}.
These challenges have further motivated innovations in advanced algorithm development, with work demonstrating that explicitly accounting for phone placement can improve activity recognition accuracy by up to 40\%~\cite{antos2014hand}, highlighting the critical importance of understanding diverse carrying behaviors for developing robust context-aware systems that function effectively across different user populations and usage contexts.

While existing research offers valuable insights into phone placement behaviors among the general population and has demonstrated the critical importance of understanding these patterns for context-aware system development, little is known about how wheelchair users may differ in their practices and preferences. Our study addresses this gap by systematically characterizing wheelchair-specific phone-carrying locations and unpacking the contextual and physical factors that influence these practices, with implications for both accessibility research and the development of inclusive context-aware technologies.

\subsection{Phone Accessibility for People with Motor Impairments}
The examination of mobile phone accessibility for people with motor impairments dates back to the early 2000s, when cellular phones began to enter mainstream consumer markets. Early work by Tomioka~\cite{tomioka2004universal} investigated the form factor of mobile phones, particularly the design of physical keypads, with participants who were blind or had upper extremity disabilities, revealing accessibility challenges related to phone size, key pitch, and key size. Kane et al.~\cite{kane2009freedom} extended this inquiry beyond lab settings through interviews and a diary study with people with vision or motor impairments, surfacing how situational factors such as lighting conditions or device use while in motion can create additional barriers. 
With the emergence of smartphones and especially touchscreens, subsequent research has largely focused on understanding usage patterns and barriers related to touch interactions in both lab~\cite{chen2013touch, duff2010effect, guerreiro2010assessing, guerreiro2010towards} and real-world settings~\cite{anthony2013analyzing, naftali2014accessibility, montague2014motor}. This has further led to a wide range of work on accessible touchscreen interaction techniques and models~\cite{mott2016smart, mott2019cluster, peng2019personaltouch}. 

Yet, most existing work has emphasized active device use, while another fundamental aspect of mobile accessibility, namely how people with motor impairments carry and retrieve their devices in everyday contexts, remains largely underexplored. Early observations emerged in Naftali et al.~\cite{naftali2014accessibility} when examining phone use experiences in mobile settings among people with motor impairments, noting that some participants placed phones on their laps or in bags attached to wheelchairs when in motion or not actively using the device. Building on these initial insights, we contribute to a more holistic understanding of phone accessibility by examining how wheelchair users carry their devices across diverse wheelchair-specific locations as well as the strategies and contextual factors that shape these practices.

\subsection{Context-aware Computing for Wheelchair Users}

Prior research on context-aware computing for wheelchair users spans three primary areas: wheelchair kinematics tracking, activity recognition, and environmental awareness. For instance, Rhodes et al.~\cite{rhodes2014validity} developed a radio-frequency-based localization system for wheelchair sports players, while Slikke et al.~\cite{van2015opportunities} and De Vries et al.~\cite{de2023real} utilized wheelchair-mounted IMUs\footnotemark \footnotetext{IMU: Inertial Measurement Unit, a sensor that detects motion by measuring acceleration and rotation.}to track kinematic features including speed, rotation, and turning magnitude. Commercial solutions have also emerged, with SMART\textsuperscript{wheel}~\cite{cooper2009smartwheel} monitoring propulsion metrics and G-WRM~\cite{hiremath2013development} using wheel-mounted gyroscopes for speed measurements.
Building upon kinematics tracking, subsequent research has advanced toward activity recognition that captures more contextually meaningful behaviors, such as daily household and desk activities~\cite{garcia2015identifying, hiremath2015detection}. More recent work has also investigated fine-grained pose estimation using either cameras~\cite{wei2016can, rammer2018assessment, milgrom2016reliability} or motion sensors~\cite{li2024wheelposer, hooke2009capturing}.
These approaches have also begun to enter consumer markets, with commercial devices like Apple Watch~\cite{WatchApp45:online} and Garmin~\cite{Wheelcha90:online} offering wheelchair-specific fitness metrics.
Complementing user-centered sensing, environmental awareness has emerged as another important area, including passive assessment of road and terrain accessibility~\cite{yairi2019estimating, iwasawa2015toward, iwasawa2016combining} and sensing sports environments through contextual cues such as dribbling sounds and game buzzers to support adaptive feedback for wheelchair athletes~\cite{carrington2020spokesense}.

Despite this growing body of work, most existing systems rely on specialized or externally mounted sensors, either on the wheelchair or the user’s body. Far less attention has been given to leveraging everyday mobile devices, such as smartphones, which offer a practical platform for scalable, real-world context sensing. Our work aims to inform the design of such systems by examining how wheelchair users physically carry their phones, a factor that significantly 
affects how sensors are situated in real-world use and, consequently, the reliability of mobile context-aware~applications.

\section{Survey on Wheelchair Users' Preferences for Phone-Carrying Locations}
To establish a broad understanding of phone-carrying location preferences among wheelchair users, we first conducted an online survey using the Qualtrics platform, collecting responses from 91 participants. To be eligible for the study, participants were required to be current wheelchair users aged 18 or older and be able to communicate in English. We did not impose exclusion criteria based on wheelchair type, diagnosis, or level of independence.
Participants were recruited via two main strategies: 1) targeted posts in wheelchair-related Reddit and Facebook groups, and 2) email and web outreach through professional networks such as the United Spinal Association~\cite{HomeUnit54:online}. 
For social media recruitment, we followed each community’s policies, including obtaining moderator approval or using designated recruitment threads where appropriate.
As compensation for their participation, participants were entered into a raffle for a \$25 Amazon gift card, with one winner randomly selected for every 25 valid responses.
The recruitment and study procedure was approved by the Institutional Review Board (IRB).

\subsection{Survey Design}
The survey included both structured and open-ended questions to gather information about participants’ wheelchair use, phone-carrying habits, and demographics. It began by asking participants to identify their primary wheelchair type (manual, power, or both). 
To explore common phone-carrying locations, participants selected up to three frequently used storage locations from a predefined list, informed by relevant online posts from wheelchair users on social media platforms and by prior work on non-wheelchair users~\cite{wiese2013phoneprioception,ichikawa2005s}. Options included trouser pockets, upper-body pockets, chest bags, underseat bags, bags attached to the back of the wheelchair, phone holders, and pockets on armbands or straps. An ``Other'' option with a text field captured additional locations.
To understand the contextual factors affecting phone-carrying decisions, we included an open-ended question prompting participants to describe situations in which they switch locations. These responses informed our interview protocol by highlighting nuanced decision-making. 
The survey also collected specific wheelchair models and demographic data (e.g., gender identity and age range) to examine correlations with phone-carrying patterns and ensure sample diversity.

\subsection{Survey Analysis}
We initially collected 693 responses and applied several filtering steps to ensure data validity. We first used Qualtrics' built-in reCAPTCHA~\cite{CaptchaV63:online} to remove responses with a score below 0.5, indicating likely bot activity ($N = 327$). We then used the RelevantID API~\cite{FraudDet65:online} to flag and remove duplicate responses originating from the same browser or device fingerprint ($N = 104$).
One researcher then manually reviewed the remaining responses and removed entries that were: 1) incomplete ($N = 7$), 2) suspicious duplicates or malicious submissions (e.g., identical metadata submitted in rapid succession; $N = 71$), or 3) inconsistent in reported wheelchair model versus selected wheelchair type, such as listing a power wheelchair model while selecting ``manual wheelchair'' ($N = 93$). This process yielded 91 valid responses, with the high removal rate reflecting common challenges with bot and spam responses in online survey recruitment~\cite{xu2022threats, sherman2024too}. All valid responses originated from within the United States.

As the survey aimed to characterize phone-carrying preferences among wheelchair users, we focused on descriptive analyses of trends across the 91 respondents. We summarized responses to multiple-choice questions by frequency and compared patterns by wheelchair type, age group, and gender. Because participants could select up to three phone-carrying locations, chi-squared tests of independence were conducted on the 156 total location selections from 91 participants rather than mutually exclusive participant-level responses. 
Due to the small sample size of dual wheelchair users ($N=6$), some expected cell frequencies fell below 5, and results involving this group should be interpreted with caution. When significant associations were found, we performed Bonferroni-corrected post-hoc pairwise comparisons~\cite{armstrong2014use} to control for multiple comparisons, with adjusted p-values reported throughout and significance evaluated at $\alpha = .05$ after correction.
Open-ended responses were analyzed using affinity diagramming~\cite{hartson2012ux} to surface initial themes behind participants’ carrying decisions.

\section{Survey Findings}
\subsection{Survey Respondents Demographic Information}
Among the valid respondents, 46 (50.5\%) identified as women, 42 (46.2\%) as men, and 3 (3.3\%) as non-binary or preferred not to disclose. Participants' ages ranged from 18 to over 55 years, with the following distribution: 18--24 (12, 13.2\%), 25--34 (31, 34.1\%), 35--44 (29, 31.9\%), 45--54 (12, 13.2\%), and 55+ (7, 7.7\%). In terms of wheelchair use, 51 (56.0\%) were power wheelchair users, 34 (37.4\%) were manual wheelchair users, and 6 (6.6\%) used both types.

\subsection{Phone-Carrying Locations Overview}
\label{survey_overview}
\begin{figure}[h]
 \centering
\includegraphics[width=0.85\textwidth]{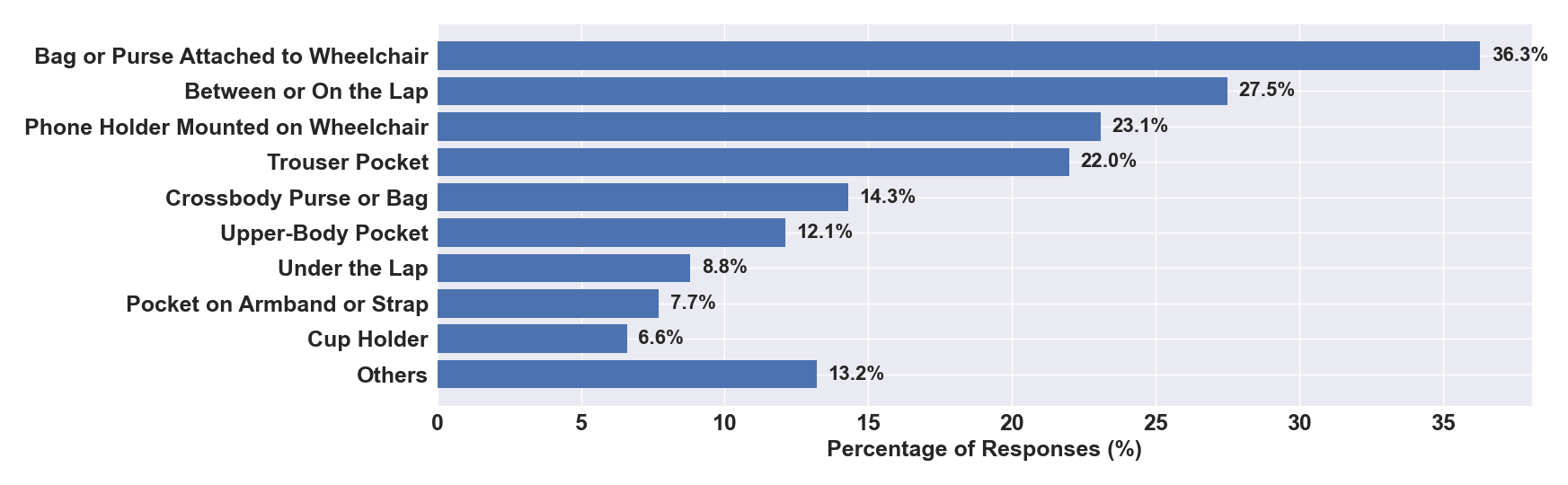}
\caption{This figure illustrates the percentage of responses for different phone-carrying locations among wheelchair users. The ``Others'' category includes wheelchair tray tables, pockets built into the wheelchair, waistbands, purses placed on laps, lanyards, and phone storage between the legs and wheelchair side guards.}
\label{fig:overview}
\end{figure}

The common locations reported by participants for storing their phones (participants could select up to three locations) showed diverse preferences, as shown in Figure~\ref{fig:overview}. 
The most frequently selected locations were bags or purses attached to the wheelchair (36.3\%), between or on the lap (27.5\%), phone holders mounted on the wheelchair (23.1\%), and trouser pockets (22.0\%). Within the ``bag or purse attached to the wheelchair'' category, participants most often placed bags under the wheelchair seat (18.7\%, \(N=17\)), followed by bags attached to the armrest or side (9.9\%, \(N=9\)) and those attached to the back of the wheelchair (7.7\%, \(N=7\)). Less frequent options included crossbody bags, upper-body pockets, storage under the lap, armband pockets, and cup holders, each selected by fewer than 15\% of participants.

Additional locations were reported by 13.2\% of participants (\(N=12\)) and grouped under ``Others'' (each under 5\%). These included wheelchair tray tables (\(N=2\)), built-in wheelchair pockets (\(N=2\)), waistband (\(N=1\)), purse or bag on the lap (\(N=1\)), lanyards hanging around the neck (\(N=3\)), and between the legs and side guards (\(N=3\)).

\subsection{Phone-Carrying Locations Across Wheelchair Types}
\label{survey_wheelchair}
We first examined whether phone-carrying preferences varied by wheelchair type, as shown in Table~\ref{tab:chair}. 
For power wheelchair users, the most common locations were bags or purses attached to the wheelchair (37.3\%, $N=19$), phone holders (33.3\%, $N=17$), and the lap area (21.6\%, $N=11$). Manual wheelchair users most frequently stored phones on or between the lap (38.2\%, $N=13$), followed by bags attached to the wheelchair and trouser pockets (each 29.4\%, $N=10$). Among users of both wheelchair types, top choices included bags or purses on the wheelchair (66.7\%, $N=4$), phone holders (50.0\%, $N=3$), and trouser pockets (50.0\%, $N=3$).

A chi-squared test of independence showed a significant association between wheelchair type and phone location ($\chi^2(18) = 38.00$, $p = .0039$, Cramér's $V = .349$). Bonferroni-corrected post-hoc comparisons revealed significant differences between power and manual wheelchair users ($\chi^2(9) = 31.83$, adjusted $p = .000638$, Cramér's $V = .475$), but not between power wheelchair users and dual users ($\chi^2(8) = 5.25$, adjusted $p = 1.000$, Cramér's $V = .238$), or manual wheelchair users and dual users ($\chi^2(9) = 15.80$, adjusted $p = .213372$, Cramér's $V = .450$). Standardized residuals~\cite{field2024discovering} indicated that power wheelchair users were more likely to use mounted phone holders ($r = 2.01$), while manual users favored placing phones under the lap ($r = 2.65$).

\begin{table}[h]
\small
\renewcommand{\arraystretch}{1}
\begin{tabular}{lccc}
\hline
\textbf{Carrying Location}                & \textbf{Power Wheelchair (51)} & \textbf{Manual Wheelchair (34)} & \textbf{Both (6)} \\ \hline
Bag or Purse Attached to Wheelchair      & \textbf{37.3\% (19)}              & 29.4\% (10)               & \textbf{66.7\% (4)}   \\
Between or On the Lap                    & 21.6\% (11)              & \textbf{38.2\% (13)}               & 16.7\% (1)   \\
Phone Holder Mounted on Wheelchair       & 33.3\% (17)\textsuperscript{$\dagger$}  & 2.9\% (1)  & 50.0\% (3)   \\
Trouser Pocket                          & 13.7\% (7)               & 29.4\% (10)               & 50.0\% (3)   \\
Crossbody Purse or Bag                   & 11.8\% (6)               & 14.7\% (5)                & 33.3\% (2)   \\
Upper-Body Pocket                       & 3.9\% (2)                & 23.5\% (8)                & 16.7\% (1)   \\
Under the Lap                    & -   & 23.5\% (8)\textsuperscript{$\dagger$}  & -            \\
Pocket on Armband or Strap            & 9.8\% (5)                & 2.9\% (1)                 & 16.7\% (1)   \\
Cup Holder                               & 7.8\% (4)                & 5.9\% (2)                 & -            \\
Others                                   & 13.7\% (7)               & 14.7\% (5)                & -            \\ \hline
\multicolumn{4}{l}{Note: Participants could select up to three phone-carrying locations, so percentages do not sum to 100\%.} \\
\multicolumn{4}{l}{Bolded values indicate the most common location for each wheelchair type.} \\
\multicolumn{4}{l}{\textsuperscript{$\dagger$} Standardized residual $|r| > 1.96$: Phone Holder, Power ($r = 2.01$); Under the Lap, Manual ($r = 2.65$).} \\
\multicolumn{4}{l}{Bonferroni-corrected post-hoc comparisons evaluated at $\alpha = .05$ (adjusted $p$-values reported in text).} \\ \hline
\end{tabular}
\caption{Phone-carrying location preferences across different wheelchair types.}
\label{tab:chair}
\end{table}

\subsection{Phone-Carrying Locations Across Gender and Age}
Drawing on prior research with the general population, which found gender- and age-specific differences in phone location preferences~\cite{ichikawa2005s,nithiya2020mobile,zeleke2022mobile}, we also examined wheelchair users' preferences across gender identities and age groups. However, our chi-squared tests revealed no significant associations between phone location and either gender ($\chi^2(18) = 14.85$, $p = .672$, Cramér's $V = .218$) or age ($\chi^2(36) = 29.64$, $p = .764$, Cramér's $V = .218$). Bonferroni-corrected post-hoc analyses likewise revealed no significant pairwise differences across gender or age groups (all adjusted $p > .05$), and standardized residuals indicated no cells with meaningful deviations from expected counts. However, confidence intervals around several subgroup proportions were wide, particularly for smaller groups, suggesting limited precision in these null findings.
As such, the detailed breakdowns are included in Appendix~\ref{survey_age_gender} (Tables~\ref{tab:gender}, \ref{tab:age}, \ref{tab:gender_ci_key}, and \ref{tab:age_ci_key}).


\subsection{Situations When Phone-Carrying Locations Change}
\label{survey_openquestion}
Beyond the commonly reported recurring carrying locations, our analysis of the open-ended responses suggested that phone placement was often context-dependent rather than fixed, offering an initial view into how carrying behavior changes across everyday situations.
First, \textbf{transitioning between environments} was a common scenario in which users adapted their phone storage strategies. These transitions included moving between home and public areas, or from familiar to unfamiliar environments.
Second, \textbf{varying road and weather conditions} also prompted frequent adjustments. Participants described modifying phone placement in response to challenging terrain, such as uneven sidewalks, ramps, or curbs. For instance, some would shift their phones from laps to a more secure location when navigating curbs. Similarly, inclement weather led users to opt for more protective placements, such as jacket pockets, to prevent phones from getting wet.
Finally, variations in phone-carrying location were also closely linked to different \textbf{activity contexts}. During tasks such as wheelchair transfers, participants preferred storage locations that reduced interference. The frequency and type of phone use further influenced placement decisions. When actively interacting with their phones, users favored easily accessible spots such as the lap or tray. In contrast, during transit or periods of non-use, phones were typically stored in bags, pockets, or other secure locations.

\section{Interview on Factors Influencing Phone-Carrying Locations}
To dive deeper into the factors shaping wheelchair users' phone-carrying behavior and preferences, we further conducted a semi-structured interview study with 15 wheelchair users. Participants were required to be current wheelchair users aged 18 or older, located in the United States, and able to communicate in English. Similar to the survey study, we did not impose exclusion criteria based on wheelchair type, diagnosis, or level of independence. Caregiver-assisted users were not excluded, though all participants who enrolled were able to independently participate in the interview. Participants were recruited through online advertising, word of mouth, and local health organizations.
All interview participants were recruited independently from the survey study with no overlap, ensuring fresh perspectives to validate and extend our survey findings. Participants who completed the interview were compensated with a \$20 Amazon gift card. The recruitment and study procedure was also approved by the Institutional Review Board (IRB). Detailed demographics of our participants are presented in Table~\ref{tab:interview_wheelchair_users}.
\begin{table}[h]
\small
\renewcommand{\arraystretch}{1.1}
\setlength{\tabcolsep}{5pt}
\begin{tabular}{lcccccc}
\hline
\textbf{ID} & \textbf{Age} & \textbf{Gender} & \textbf{Wheelchair Type} & \textbf{Power Assist} & \textbf{Medical Condition} & \textbf{Years of Use} \\ \hline
W1  & 31 & Male   & Power  & N/A  & Cervical spinal cord injury     & 12  \\
W2  & 24 & Female & Power  & N/A  & Spinal muscular atrophy        & 17  \\
W3  & 47 & Male   & Manual & No   & Spinal cord injury             & 2   \\
W4  & 51 & Male   & Power  & N/A  & Muscular dystrophy             & 15  \\
W5  & 42 & Female & Power  & N/A  & C5 spinal cord injury          & 22  \\
W6  & 46 & Male   & Manual & No   & Lumbar spinal cord injury      & 13  \\
W7  & 67 & Female & Both   & No   & Spina bifida                   & 45  \\
W8  & 43 & Female & Both   & No   & Ehlers-Danlos syndrome         & 8   \\
W9  & 47 & Female & Manual & Yes  & T9 spinal cord injury          & 25  \\
W10 & 42 & Female & Power  & N/A  & Arthrogryposis                 & 39  \\
W11 & 40 & Female & Manual & Yes  & Multiple sclerosis             & 15  \\
W12 & 38 & Male   & Power  & N/A  & C6-7 spinal cord injury        & 28  \\
W13 & 26 & Female & Manual & Yes  & Fibromyalgia                   & 5  \\
W14 & 40 & Male   & Manual & No   & T5-6 spinal cord injury        & 19  \\
W15 & 23 & Female & Power  & N/A  & Spinal muscular atrophy        & 19  \\ \hline
\multicolumn{7}{l}{Note: A power assist device is a motorized accessory that can be coupled to a manual wheelchair.}\\ \hline
\end{tabular}
\caption{Demographics of wheelchair users in the interview study.}
\label{tab:interview_wheelchair_users}
\end{table}

\subsection{Interview Procedure}
These semi-structured interviews were one hour long and followed the format described below.

\subsubsection{Background ($\sim$5 minutes)} This section covered demographic information about wheelchair users, their diagnosed medical conditions, and their wheelchair usage history.

\subsubsection{Commonly Used Phone-Carrying Locations ($\sim$10 minutes)} We asked wheelchair users about their commonly used phone-carrying locations and verified the validity of the possible locations identified in our survey study.

\subsubsection{Factors Influencing Phone-Carrying Location Choices ($\sim$30 minutes)} We asked wheelchair users about the factors influencing their phone-carrying behavior and the situations in which they switched between different carrying options. Informed by our survey study, we prepared probing questions focused on wheelchair design, environmental factors (e.g., weather and terrain conditions), and activity context (e.g., guiding us through a typical day while reflecting on phone~locations).

\subsubsection{Potentials of Phone-Carrying Location Awareness ($\sim$15 minutes)}
In this section, we first explored the challenges that wheelchair users currently face with phone usage. We then asked them to brainstorm potential applications and benefits of leveraging phone-carrying location awareness to enhance usability and accessibility.

\subsection{Interview Analysis}
The semi-structured interviews were conducted via Zoom~\cite{zoom}, and participants were invited to use any assistive technology or accommodations they needed; no additional accommodations were requested. Interviews were audio-recorded with participants' consent and then transcribed, yielding 13.6 hours of interview data.
For analysis, we applied Braun and Clarke’s reflexive thematic analysis approach~\cite{braun2006using} to guide an iterative and interpretive analysis of participants’ experiences. 

Specifically, we collated the transcripts and notes taken during the interviews, and the first author conducted initial open coding, generating 221 codes grounded in participants’ accounts. The research team then met regularly over the course of a month to collectively examine the evolving codes, discuss interpretations, and reflect on emerging patterns. Disagreements in coding or interpretation were surfaced during these meetings and resolved through discussion and consensus-building, with particular attention to revisiting the original data to ensure that interpretations remained grounded in participants’ accounts~\cite{braun2019reflecting}.
Based on these discussions, the first author iteratively refined the codes between meetings.
Once all codes reached shared agreement, the research team collectively organized the finalized codes into thematic groups through affinity diagramming~\cite{hartson2012ux} on a Miro board~\cite{miro}. Over an additional month of continued iterative meetings, the team reviewed and refined the thematic structure, clarified relationships between themes, and developed the final set of themes presented in this paper, each of which corresponds to one subsection in Sec.~\ref{interview_results}. We assessed theme adequacy based on the richness, coherence, and explanatory power of the final thematic structure rather than a predetermined saturation threshold, consistent with reflexive thematic analysis guidance~\cite{braun2021can}.

\subsection{Positionality Statement} In line with reflexive practices, we acknowledge how our experiences and positionality shape our perspectives and this analysis. We are both researchers in Human-Computer Interaction working in the United States. Collectively, the team has over a decade of experience conducting accessibility research with wheelchair users, ranging from building novel mobile and wearable sensing systems to conducting qualitative and mixed-methods user research on understanding everyday accessibility challenges. Neither author is a wheelchair user. To conduct this research, we engaged directly with wheelchair user communities through online platforms, advocacy organizations, and local health networks, and sought to center participants' own language and priorities throughout our analysis.

\section{Interview Findings}
\label{interview_results}
Here, we present our findings across six subsections, each corresponding to a theme that emerged from our analysis. We begin by outlining participants’ general adoption patterns in phone-carrying behaviors. We then discuss how phone-carrying locations are typically first guided by considerations of reachability, which are often grounded in each individual's physical abilities and the design of their wheelchair. After that, we examine two additional factors, safety and availability, that often influence how participants choose or alternate between these reachable locations. Finally, we describe two contextual factors that also affect in-situ decisions, namely the social environment and the use of other devices such as smartwatches.
 
As a starting note, we validated the phone-carrying locations identified in our survey study with our interview participants. All reported locations were confirmed by participants as either ones they personally used or had observed being used by other wheelchair users.

\subsection{Adoption}

Extending the survey findings (Sec.~\ref{survey_overview}), the interview participants also demonstrated diverse practices and preferences in their phone-carrying locations.
These ranged from common storage options such as trouser or upper-body pockets, to around-body placements like on or under the lap, between the legs, or tucked between the body and wheelchair components. Many participants also relied on wheelchair-mounted solutions, including underseat bags, armrest-mounted purses, phone holders, cup holders, tray tables, and lanyards hung on brake handles or joysticks. A few further experimented with wearable accessories, such as straps or bands secured to their arms or legs, underscoring the high degree of personalization. 

A key reason for this variability lies in the different needs, physical abilities, and wheelchair configurations of each individual. As W11 noted:
\textit{``Everybody is different, and everybody’s chairs are different, so what works for me and my chair won’t necessarily work for someone else. We’re all different in terms of what’s helpful. Everyone does it their own way, and in the end, you just make it work because you have to.''}

Much like the adoption and customization of other assistive technologies~\cite{hurst2011empowering}, most participants arrived at their preferred phone-carrying solutions through \textbf{trial and error}. Over time, they developed habitual approaches tailored to their specific needs and contexts. Rather than relying on prescribed methods, participants often engaged in considerable experimentation, as W6 explained: \textit{``It kind of evolved \dots{} when I first got the chair, I didn't have the backpack, and I got the backpack, and then I had a little fanny pack on the front, but I had to unzip it to get it out which wasn't that convenient \dots{} now I switched to use this pocket in my seat cushion.''}

Once wheelchair users establish a routine that works for them, they tend to adhere to it rather than experiment with new alternatives. 
This preference for consistency often persists even when users transition to a different wheelchair (W2, W5, W7, W9, W10-W12, W14, W15). For example, W14 noted that: \textit{``All my six chairs have had that underseat pouch, and for 19 years, the phone has been there.''}
Likewise, W10 described proactively ensuring that her new chair accommodated her existing phone-carrying setup:
\textit{``I even had to tell them, 'I need a side piece here because I need to be able to put things there.' That’s just become part of my needs.''}

\subsection{Reachability}
Regarding the actual decision-making process of choosing potential phone-carrying locations, reachability emerged as the most critical factor influencing where wheelchair users \textbf{can} reliably place and retrieve their phones. Moreover, reachability is often determined by each wheelchair user's physical abilities and the physical design of their wheelchair.

Specifically, many participants emphasized that their phone-carrying locations are tied to their \textbf{range of motion} (W1, W2, W10-W12, W15). For instance, W10 articulated why she keeps her phone in a bag on her left side: \textit{``I can't reach down below by my footrest. I can't reach in the backpack. Wouldn't be able to reach my right side either \dots{}.''} Similarly, W2 described why she uses a phone holder mounted near her wheelchair joystick:  
\textit{``I can't lift up my arms \dots{} my joystick is kind of the only place I can reach independently.''}   
\textbf{Hand dexterity} was also a common physical factor influencing phone placement (W1, W2, W5, W8, W12). Several participants noted that a lack of dexterity made certain locations, such as under the lap or bags with zippers, difficult or inaccessible to use, even if they seemed safer (W1, W2, W8). 

Beyond individual physical abilities, participants also reported that the \textbf{physical design of the wheelchair} significantly impacts the reachability of certain phone-carrying options. Echoing the differences identified in our survey results (Sec.~\ref{survey_wheelchair}), many participants noted that the physical design of power wheelchairs makes it much easier to attach accessories within easy reach compared to manual wheelchairs (W4, W5, W7, W8, W10, W12, W15). 
Meanwhile, certain features of power wheelchairs can also limit storage or reach. For instance, the presence of motors and battery packs can make underseat storage options unfeasible. 
Taller backrests, headrests, and larger armrests can also obstruct access to areas that are reachable in manual chairs, as W7 explained: 
\textit{``This [power wheelchair's backrest] is pretty high. I've got a headrest and the side thing [armrest]. It's hard to reach your arm around the back. When you're in a manual chair that has a lower back, you could put something on and reach it, or have the thing in the front, behind your legs [underseat bags].''}

\subsection{Availability}
Among reachable phone-carrying locations, wheelchair users further evaluate options based on device availability, defined as how readily users can access, monitor, and interact with their phone from a given location. In practice, availability reflected factors like ease of retrieval and glanceability, and was often considered alongside the intensity of phone use and mode of interaction.

Specifically, consistent with our survey findings (Sec.~\ref{survey_openquestion}), all participants mentioned adjusting their phone-carrying location based on \textbf{the intensity of use}. During continuous or active use, the most common strategy was to hold the phone in hand. Some participants also relied on a tray table or mounted phone holder, particularly those with limited grip strength (W1, W2, W5, W8, W12). When the phone was needed only intermittently or for short-term storage, users preferred locations that allowed quick access. 
Common choices for temporary storage included users’ lap (W1-W6, W8-W11, W13-W15), pockets (W9, W13, W14), cup holders (W8, W9, W14, W15), or nearby surfaces, as W14 described:
\textit{``I’ll put it between my legs if I’m just rolling around the house and texting back and forth with someone, or I’ll keep it in my hoodie pocket, on my lap, mainly because I’m in the middle of using it.''}
 In addition to how frequently the phone is used, the \textbf{mode of interaction}, whether through touch or voice, influenced where participants placed their phones (W2, W3, W5, W12, W14). For example, W5 explained:
\textit{``If I'm on social media or sending a text message, it'll be in my hands. If I asked [Siri] to give me directions somewhere, it would be on my joystick, because I won't necessarily need to hold it.''}

Another general consideration beyond active usage is \textbf{glanceability}, or how easily users can check or monitor their phones without having to retrieve them. Common locations that support glanceability include phone holders, lap placements, and tray tables.
For instance, W9 explained how she uses a cup holder as a phone mount for passive information access while her hands are busy with manual propulsion:
\textit{``I have a cup holder attached to the side of my chair and I wedge my phone in there. It's tilted up toward me, so I can just glance down to check it. That way, I don’t have to actually pick it up and use my hands, since I might be pushing my chair or wearing gloves.''}



\subsection{Safety}
Alongside availability, safety concerns also shape how wheelchair users decide where to store their phones. As they engage in various daily activities or move through different environments, wheelchair users often balance the convenience of access with the need to keep their phones secure from potential loss, damage, or theft.

First, safety concerns are particularly salient when wheelchair users anticipate \textbf{significant movement}, including bodily motions or traveling across places. Since wheelchair mobility often relies on the user’s upper body, through manual propulsion or joystick control, holding or securing a phone by hand while moving around is typically impractical. Similarly, although placing the phone on the lap allows for quick access and easy glancing, many participants expressed concerns about their phones falling while navigating different settings.
To address these concerns, wheelchair users employed a variety of strategies that provided more stability while keeping the phone reasonably accessible.
Some participants mentioned tucking the phone under their legs for added stability (W2, W3, W10), while others placed it between their legs or between the leg and the wheelchair side guards (W4, W8, W9, W11, W14, W15). Embracing a do-it-yourself (DIY) approach, W4 and W8 further described designing and 3D-printing customized phone holders to address both ease of access and safety needs. Additionally, W4 repurposed an armband originally made for joggers, adapting it into a thigh-mounted holder.

Several environmental factors were also found to be intertwined with the above safety-related considerations. \textbf{Weather conditions}, such as rain or snow, prompted participants to seek more protective phone placements (W1-W3, W5-W10, W12-W15). 
Similarly, \textbf{terrain and road conditions}, such as bumps, uneven sidewalks, and ramps, influenced placement decisions (W1, W2, W5, W6, W10-W15). As environmental hazards increased, such as when transitioning from smooth indoor flooring to rough outdoor surfaces, participants reported shifting away from more exposed placements, like their laps, toward more secure options.
As W9 emphasized:
\textit{``The more I'm outside, the more I use the phone holder, because not all concrete is smooth.''}
Further, in \textbf{crowded or public environments}, many were cautious about leaving their phones in open or visible spots. Instead, they preferred enclosed or body-adjacent storage options, such as bags, pockets, or underseat bags, to reduce the risk of theft or accidental loss.

\subsection{Use of Smartwatches}
The adoption of smartwatches also influences the way wheelchair users interact with and carry their phones, offering a practical alternative to phones in various scenarios. Since smartwatches are typically worn on the wrist, they are often in a more consistently reachable location compared to phones, which must be carried separately. As a result, many participants mentioned that smartwatches often complemented and, in some cases, reshaped their phone-carrying behaviors. 
Specifically, participants highlighted the everyday convenience of smartwatches, noting that wearable access to notifications and phone calls allowed them to store their phones in more secure, less accessible locations, such as bags or pockets, without compromising functionality (W1, W3, W8, W10, W11, W13). A few of them also emphasized that smartwatches provided an added layer of personal safety, especially in situations where phone reachability was compromised (W1, W10, W11). For instance, W11 explained: \textit{``I now have a smartwatch, so when I cannot get to my phone, I can use my watch to call when something is going terribly wrong. That’s why I bought the smartwatch in the first place.''} 

\subsection{Social Context}
Finally, social context was found to influence how wheelchair users decide where to carry their phones. One common theme was the role of other people, particularly caregivers or companions, in helping with phone storage and retrieval. 
When support was available, users often felt less constrained in their choices. For example, participants reported asking their caregivers to help store their phones in back-mounted bags that were out of reach or simply carry the phones for them for safety reasons. They also felt more comfortable keeping their phones in accessible locations, such as on their laps or in cup holders, knowing assistance was available if dropped (W1, W2, W5, W10, W12, W15). Conversely, when alone, participants were more deliberate about phone placement to ensure both independence and phone safety, as W15 shared: \textit{``It also depends on if I'm alone or with other people. Because if I'm alone and I drop the phone, I'm out of options. But if I'm with other people, someone could pick it up, so it’s not as big of a deal.''}
In addition to practical concerns, a few participants also reported making adjustments based on social norms (W1, W12, W14). For instance, during social events or conversations, some users chose to move their phones from visible or informal locations, such as their laps, to more discreet places like pockets or bags.

\section{Discussion and Future Directions}
Our investigation of phone-carrying practices among wheelchair users reveals both the complexity of mobile accessibility and new directions for inclusive technology design. In this section, we first characterize how these practices constitute a distinct form of mobile accessibility shaped by physical ability, wheelchair design, and everyday context. We then synthesize a set of design implications and future directions for wheelchair user–facing, context-aware systems, grounding each in observed carrying practices and illustrating it with concrete examples.

\subsection{Characterizing Phone Accessibility Through Carrying Practices}
\begin{figure}[h]
\centering
\includegraphics[width=0.9\linewidth]{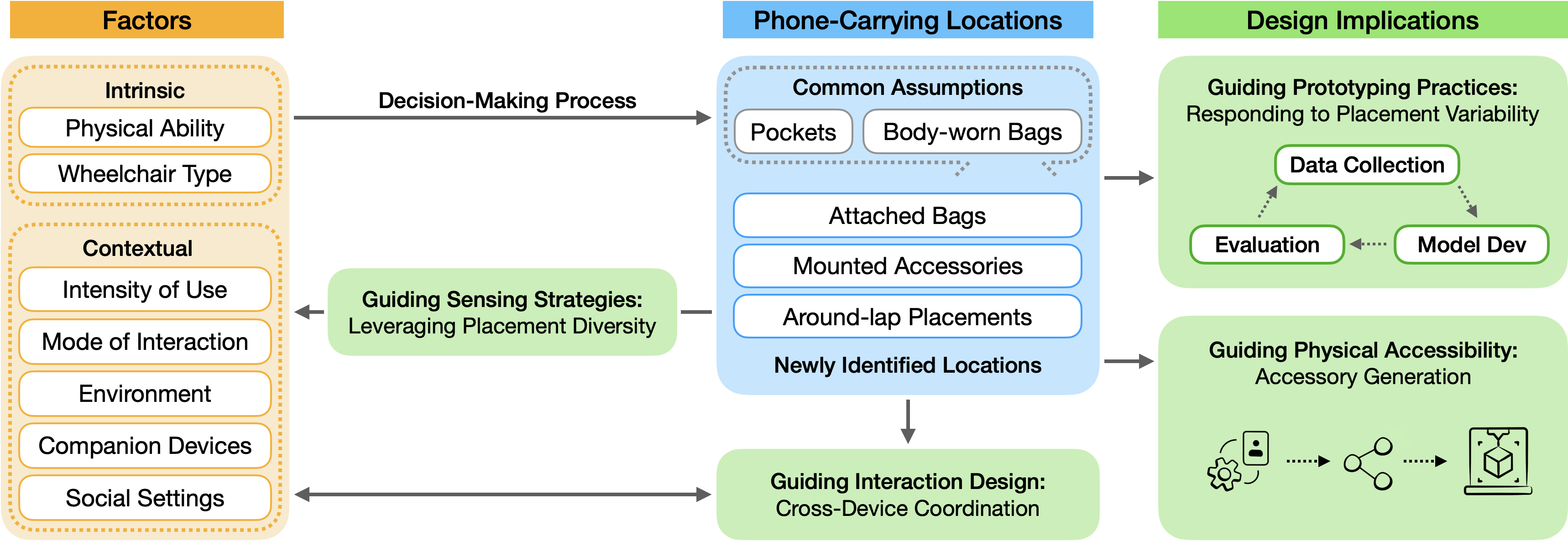}
\caption{A design framework linking factors, phone-carrying locations, and design implications. In this work, we identify wheelchair-specific phone-carrying locations that emerge from users’ decision-making processes shaped by intrinsic and contextual factors. Building on these findings, we synthesize four design implications that offer actionable guidance for inclusive prototyping, sensing, interaction, and physical accessibility.}
\Description{This figure shows a design framework that explains how different factors influence where wheelchair users place their phones and how these insights inform design implications. On the left, intrinsic factors such as physical ability and wheelchair type, together with contextual factors such as activity intensity, interaction needs, environment, companion devices, and social setting, contribute to users’ placement decisions. In the center, these decisions lead to a range of phone carrying locations, including pocket, body worn bags, lap related placements, wheelchair attached bags, and mounted holders. On the right, the framework presents four areas of design implications: prototyping, sensing, interaction, and physical accessibility. Arrows illustrate how factors shape placement and how placement guides opportunities for more inclusive mobile design.}
\label{fig:design_framework}
\end{figure} 

While the phone-carrying habits of the general population have been extensively studied and documented~\cite{wiese2013phoneprioception, redmayne2017s, dey2011getting, ichikawa2005s, zeleke2022mobile, sorysz2023beyond, nithiya2020mobile}, our study uncovers distinct and adaptive practices among wheelchair users that extend beyond conventional patterns. These practices are shaped by the interplay among physical ability, wheelchair design, and everyday contexts, enriching our understanding of mobile accessibility beyond active interaction. 

In contrast to prior findings ~\cite{ichikawa2005s, redmayne2017s, zeleke2022mobile}, our analysis revealed no significant associations between phone-carrying preferences and age or gender. Instead, \textbf{physical ability and wheelchair design were the most salient intrinsic factors}, underscoring that for wheelchair users, carrying practices are likely less about lifestyle preferences and more about accessibility and functional needs. The substantial variability in both individual physical capacities (e.g., range of motion, strength, dexterity) and wheelchair configurations (e.g., power vs. manual chairs, custom features) means that phone-carrying practices emerge from the tight coupling of body and mobility aids, making them inherently personalized.

This personalization further manifests through a continuous process of experimentation and refinement, as users seek to balance independence, device usability, and safety. Participants described evaluating potential storage locations primarily in terms of reachability, ensuring the phone could be independently and reliably accessed, and then weighing availability against safety across situations. Through repeated daily experience, they developed candidate locations and dynamically switched between options based on activity, routine, terrain, weather, social setting, and the use of companion devices. This ongoing negotiation suggests that phone carrying is not a fixed behavior but an evolving accessibility practice embedded within everyday mobility.

Together, these findings suggest that phone carrying is not merely a logistical choice, but a form of accessibility work through which wheelchair users adapt mobile technology to their bodies, chairs, and environments. Importantly, because these practices emerge in response to users’ embodied needs and situational constraints, they also constitute a valuable resource for mobile context-awareness. Figure~\ref{fig:design_framework} maps this relationship by showing how intrinsic and contextual factors shape carrying locations, which in turn inform design implications across four directions: interaction and cross-device coordination, where placement can drive system adaptations; physical accessory design and generation, where carrying practices surface unmet accessibility needs; prototyping workflows, which must account for placement variability to ensure application effectiveness; and sensing, where diverse placements open new opportunities for capturing user, wheelchair, and environmental context. Below, we begin with interaction and physical accessibility, which draw directly on participants' reported experiences, then turn to prototyping and sensing, where preliminary data explorations reinforce necessity and illustrate feasibility, respectively. We close with ethical considerations that cut across these directions. The examples throughout are intended to be suggestive rather than exhaustive, highlighting design opportunities and open directions for the broader community.

\subsection{Interaction: Placement-Aware and Cross-Device Coordination}

\subsubsection{Implication}
\label{dis:placement_as_context_sub}
Our findings show that wheelchair users’ phone-carrying choices are often deliberate and context-sensitive, encoding meaningful information about intent, attention, and mobility. As such, \textbf{phone placement can serve as a proxy for user context} that context-aware systems can leverage to adapt interaction and feedback. For example, a phone placed on a lap, tray, or mounted holder may indicate readiness for active engagement, whereas a phone stored in an underseat bag or backpack often signals that the user is prioritizing mobility or safety.

At the same time, participants’ accounts highlight a recurring tension between availability and safety: phones are often placed out of reach during propulsion or navigation, with smartwatches helping maintain lightweight access. This suggests that accessibility challenges arise not only from interface design but also from reachability constraints and mobility demands embedded in everyday use. Extending prior work on situational impairments in mobile interaction~\cite{kane2009freedom, naftali2014accessibility, wobbrock2019situationally}, our findings point to an interaction paradigm that is both placement-aware and ecosystem-oriented. In this paradigm, systems should adapt interaction to where the phone is and leverage cross-device coordination when direct interaction is impractical~\cite{kubo2017exploring, brudy2019cross}.

\subsubsection{Next steps}
Building on this implication, designers should move beyond phone-centric experiences toward systems that dynamically distribute sensing and interaction across devices. This includes taking advantage of the growing adoption of smartwatches, the expansion of smart wheelchair features, and the emergence of new wearables such as smart glasses. Existing cross-device frameworks, such as Brudy et al.’s taxonomy~\cite{brudy2019cross}, offer transferable scaffolds for coordinating interaction across configuration, engagement, and disengagement phases. Within this framing, systems can assess reachability, route interaction to a watch when the phone is stowed, delegate sensing to wheelchair-mounted phones, and return control once the phone becomes reachable. Complementary design spaces and toolkits for smartphone–smartwatch interaction~\cite{kubo2017exploring, houben2015watchconnect} provide concrete techniques for implementing such flows.

Critically, however, current cross-device frameworks rarely treat the \textbf{wheelchair as a first-class device}. While the chairables lens~\cite{carrington2014wearables} argues that wheelchairs can host input, output, and sensing capabilities, this perspective has not been fully integrated into mainstream cross-device design guidelines. Our findings provide empirical grounding for extending these frameworks, showing that offloading interaction or sensing to the wheelchair becomes particularly valuable when phones are deliberately placed out of reach for safety or mobility reasons. Power wheelchairs, with built-in control panels and computing capabilities, offer immediate opportunities for such integration. Manual wheelchairs, by contrast, present fewer built-in affordances, but designers could repurpose components such as pushrims, armrests, seat cushions, or backrests as complementary input or output channels. Future work should explore \textbf{chairables-integrated cross-device toolkits} for experience prototyping and evaluate their effectiveness in supporting the dynamic, context-sensitive interaction strategies wheelchair users already employ.

\subsubsection{Illustrative examples}
One illustrative application of placement-aware, cross-device interaction is adaptive notification and input design. Building on the principle that phone placement encodes user context, recognizing placement enables feedback strategies that reduce effort and missed events. Participants recommended prioritizing vibration for lap placements and audio for placements such as bags or holders (W2, W4, W7, W8, W10, W13). Beyond fixed defaults, systems could offer customizable profiles, similar to Focus or Sleep mode, that allow wheelchair users to define notification strategies for specific placements and routines.

Placement recognition can also inform adaptive input modalities. When phones are secured in exposed positions such as mounted holders, systems may prompt hands-free interaction options, including voice input, to reduce interaction effort~\cite{wiese2013phoneprioception} (W2, W4, W10, W12). During wheelchair movement, dynamic touch sensitivity adjustments could mitigate vibration-induced errors, drawing on techniques developed for touch interaction while walking~\cite{mott2019cluster}. Conversely, when phones are stored under or between the legs, systems could detect and suppress unintended touch input, reducing accidental interactions reported by participants (W4, W8).

\subsection{Physical Accessibility: Context-Aware Accessory Design}

\subsubsection{Implication}
Our findings show that wheelchair users frequently repurpose existing accessories (e.g., cup holders, side guards, seatbelts) and experiment with DIY mounting solutions to achieve workable phone placements. These practices reveal both unmet design needs and strong user agency in creating personalized configurations that balance reachability, safety, and stability. Rather than treating such adaptations as temporary workarounds, they should be understood as \textbf{actionable design signals} that point toward context-aware approaches to physical accessory design. In this framing, physical accessibility is not a fixed property of a device or mount, but an outcome of how accessories are tailored to individual wheelchair configurations, activities, and reachability constraints.

\subsubsection{Next steps}
Building on this implication, future work should explore tools and workflows that lower barriers to creating personalized, durable mounting solutions by translating user context into explicit design requirements. Relevant contextual factors include wheelchair model and geometry, activity context, phone placement preferences, and individual reachability constraints. Participants’ challenges around fabrication skills, design file creation, and material durability mirror broader limitations documented in personal fabrication and DIY assistive technologies~\cite{mahapatra2019barriers, hudson2016understanding, mcdonald2016uncovering, li2024exploring, hook2014study}. Addressing these barriers requires design support that bridges everyday use contexts and fabrication expertise, rather than assuming technical proficiency or access to specialized tools. 

\subsubsection{Illustrative examples}
One illustrative direction is the use of AI-assisted generative design tools to support context-aware accessory creation. For instance, users could upload images of their wheelchairs and describe desired functionality, while systems combining image-to-3D capabilities~\cite{lai2025hunyuan3d25highfidelity3d} with large language models interpret design constraints and generate customizable 3D-printable files. In parallel, designers and organizations could \textbf{maintain curated repositories} in which users, clinicians, and makers share parametric templates, accessory profiles, and best practices. Such repositories could enable contextual recommendations based on wheelchair configuration, activity context, and user ability, supporting more sustainable and accessible personalization. Together, these approaches point toward an ecosystem that combines intelligent tooling with shared community knowledge to improve physical phone accessibility in everyday wheelchair use.

\subsection{Prototyping: Responding to Placement Variability} 
\subsubsection{Implication}
\begin{figure}[h]
\centering
\includegraphics[width=\linewidth]{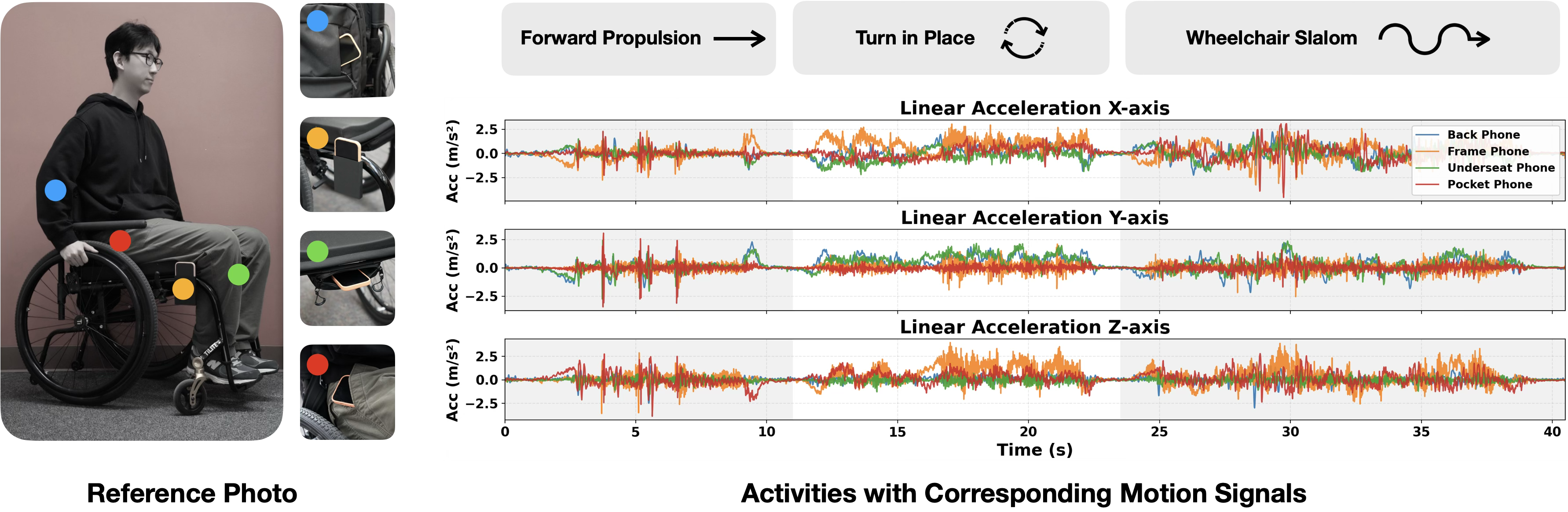}
\caption{Signal divergence across four carrying locations. Synchronized accelerometer traces (X, Y, Z) from phones in a back bag, frame holder, underseat bag, and pocket during (1) propulsion, (2) turning, and (3) slalom. The distinct waveforms and low correlation coefficients (Table~\ref{tab:correlation}) confirm that identical activities yield unique sensor signatures depending on placement.}
\Description{A composite figure. On the left, a photo of a wheelchair user with green dots indicating phone locations: arm, pocket, frame, and underseat. Beside it, close-ups of the phone holders. To the right, three line graphs show linear acceleration for X, Y, and Z axes. The lines are colored differently for each phone location and show significant variation. Above, a legend identifies these as Forward Propulsion, Turn in place, and Slalom.}
\label{fig:signal_variation}
\end{figure}

Designers and developers should move beyond assumptions that phones are typically held in hand or kept in pockets when developing mobile applications for wheelchair users. This is particularly important for context-aware systems, as phone placement directly affects both the quality and interpretation of sensed data~\cite{coskun2015phone, henpraserttae2011accurate}. Figure~\ref{fig:signal_variation} grounds this relationship empirically by showing accelerometer traces from identical manual wheelchair activities (propulsion, turning in place, and slalom) captured simultaneously across four carrying locations: trouser pocket, underseat bag, back-attached bag, and frame-mounted holder. Even for these basic wheelchair maneuvers, the signals vary substantially in amplitude, frequency characteristics, and noise patterns, as reflected in the consistently low cross-location correlation coefficients in Table~\ref{tab:correlation}. These variations directly affect downstream feature extraction and classification performance, reinforcing that understanding wheelchair users' actual carrying practices is essential for building robust context-aware systems.

\begin{table}[h]
\small
\renewcommand{\arraystretch}{1}
\begin{tabular}{cccc}
\hline
\textbf{Axis} & \textbf{Forward Propulsion} & \textbf{Turn in Place} & \textbf{Wheelchair Slalom} \\ \hline
X & $-0.04 \pm 0.39$ & $0.07 \pm 0.34$ & $0.06 \pm 0.28$ \\
Y & $0.18 \pm 0.27$  & $0.11 \pm 0.22$ & $0.11 \pm 0.16$ \\
Z & $0.06 \pm 0.13$  & $0.09 \pm 0.12$ & $0.08 \pm 0.13$ \\ \hline
\end{tabular}
\caption{Correlation coefficients (0 = none, 1 = perfect) for accelerometer axes across carrying locations during identical wheelchair activities. Low values confirm substantial placement-induced differences in sensed~motion.}

\label{tab:correlation}
\end{table}

\subsubsection{Next steps}
To build inclusive and effective applications, designers should incorporate real-world carrying practices into data collection, model training, and system evaluation. This could involve: (1) capturing placement labels during collection and publishing the label schema and protocol for reproducibility, and (2) stratifying benchmarks to report system performance per placement as well as under realistic placement mixtures. Furthermore, given that users dynamically switch between locations, context-aware systems must be robust to these changes. This requires adaptive algorithms that either explicitly detect and account for carrying location or function effectively regardless of placement. Developers may treat placement as a domain variable, comparing domain-adaptation approaches against ensembles of placement-specialized models routed by a lightweight classifier. Additional strategies include placement-aware data augmentation (e.g., simulating bag occlusion or holder vibration), contrastive learning~\cite{haresamudram2025past} for placement-invariant representations, and few-shot personalization~\cite{xu2022enabling} to new chair--user configurations.

\begin{figure}[H]
\centering
\includegraphics[width=0.9\linewidth]{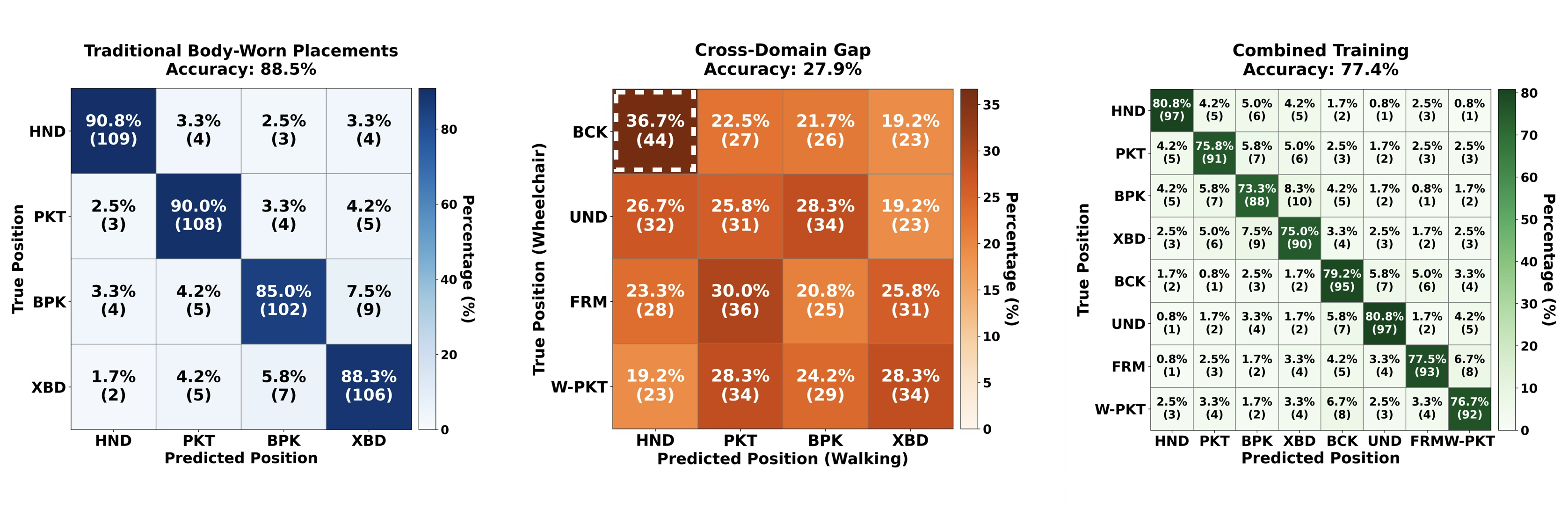}
\caption{Preliminary phone-carrying location recognition across walking and wheelchair contexts. Walking locations: hand (HND), trouser pocket (PKT), backpack (BPK), and crossbody bag (XBD). Wheelchair locations: back-attached bag (BCK), underseat bag (UND), frame-mounted holder (FRM), and trouser pocket (W\text{-}PKT).
Left: A classifier trained on traditional body-worn locations performs well within-domain. Middle: When tested on wheelchair-specific locations, the walking-trained model exhibits a pronounced cross-domain gap (accuracy = 27.9\%), with the highlighted top-left cell illustrating a representative error where phone-in-bag samples are misclassified as hand-held. Right: Training on combined walking and wheelchair data improves recognition across all locations.}
\label{fig:location_recog}
\end{figure}

\subsubsection{Illustrative example: Inclusive phone-carrying location recognition}
Prior research has shown that smartphone sensors can be used to detect phone-carrying locations in the general population~\cite{wiese2013phoneprioception, coskun2015phone, sztyler2016body, gao2019context}. However, it is unclear whether these methods generalize to the wheelchair-specific positions identified in our study. To probe this, we conducted a preliminary exploration in which the first author collected IMU data at 100~Hz using iOS's Core Motion API during 12 minutes of walking and 12 minutes of wheelchair locomotion, with four phones (iPhone 14 Pro) carried simultaneously in different locations. Walking placements included hand, trouser pocket, backpack, and crossbody bag, while wheelchair placements included back-attached bag, underseat bag, frame-mounted holder, and trouser pocket.
Following established methods~\cite{wiese2013phoneprioception, coskun2015phone, sztyler2016body}, we extracted accelerometer and gyroscope features over 3-second non-overlapping windows, including mean, variance, root mean square, interquartile range, and the 25th, 50th, and 75th percentiles for each axis. We then divided each dataset into 6 minutes for training and 6 minutes for testing, and trained an SVM classifier. As shown in Figure~\ref{fig:location_recog}, the classifier trained solely on ambulatory data, reflecting traditional body-worn assumptions, achieved high accuracy within the walking dataset, consistent with prior work. However, when applied to wheelchair data, it struggled to separate even coarse categories (e.g., pockets vs.\ bags). Retraining on combined walking and wheelchair data substantially improved performance.
These challenges likely arise from wheelchair-specific motion characteristics and further highlight the importance of correctly accounting for wheelchair users’ phone-carrying behavior during model development. While this preliminary exploration involved a single user and is intended only as an initial proof of concept, it highlights the need for broader validation with wheelchair users across diverse carrying locations and mobility contexts. Future work should evaluate such models using subject-independent protocols such as leave-one-subject-out cross-validation, compare performance across realistic placement mixtures, and test robustness to transitions between placements during everyday mobility.

\subsection{Sensing: Leveraging Placement Diversity}
\label{dis:placement_as_context}

\subsubsection{Implication}
Our findings suggest that the diversity of phone-carrying practices among wheelchair users is not only a challenge for prototyping but also a resource for mobile sensing. Specifically, we argue that authentic phone carrying practices, spanning wheelchair-mounted accessories and around-body placements, (1) enable practical replication of sensing solutions that previously relied on external sensors, and (2) create novel sensing and context-awareness opportunities that differ from common pocket or handheld assumptions. Central to this implication is the idea of \textbf{opportunistic sensing}. Rather than prescribing fixed device placements, systems can leverage how phones are naturally carried and adapt sensing strategies as users transition between locations. In this framing, placement is not noise to be corrected but a contextual structure that reflects how sensing can be integrated into everyday mobility.

\subsubsection{Next steps}
Building on this implication, future mobile sensing systems for wheelchair users could opportunistically select sensing modalities and inference strategies based on current phone placement. This includes leveraging body-coupled placements for capturing user motion and effort, chair-mounted placements for capturing wheelchair dynamics and environmental conditions, and transitions between placements as signals of changing intent or activity. Over longer time scales, longitudinal placement patterns may further reveal routines and activity transitions, enabling anticipatory and adaptive sensing behaviors tailored to wheelchair users’ everyday mobility.

\subsubsection{Illustrative examples}
\begin{enumerate}

\item \textbf{Chair-mounted placements for wheelchair kinematics and condition monitoring.} Chair-mounted placements such as holders, tray tables, or underseat bags mechanically couple the phone to the wheelchair frame, yielding motion signals that primarily reflect wheelchair dynamics. These placements enable sensing capabilities traditionally achieved through external instrumentation (e.g., wheel-mounted IMUs), including tracking speed, trajectory, and orientation~\cite{rhodes2014validity, carrington2020spokesense, van2015opportunities}, using built-in smartphone sensors. Over longer periods, gradual changes in vibration spectra or audio signatures may also indicate wheelchair condition issues such as tire wear, caster misalignment, or frame looseness, supporting proactive maintenance needs that wheelchair users have identified as critical~\cite{mo2024exploring, hogaboom2018wheelchair}. Future work should validate this opportunity with multi-participant studies that compare phone-based estimates from chair-mounted placements against established ground-truth systems, such as wheel-mounted encoders or motion capture. Evaluation should examine both within-subject and subject-independent performance and test generalization across wheelchair types, mounting configurations, and everyday routes. Longer-term deployments could further assess whether changes in vibration or acoustic features reliably correspond to real maintenance events or component degradation over time.

\item \textbf{Body-coupled placements for activity recognition and mobility profiling.} Meanwhile, body-coupled phone placements, such as pockets or positions under the lap, expose motion signals that reflect users’ movement dynamics, enabling ``wheelchair user modes''\footnote{For example, Apple Watch and Garmin devices offer wheelchair-specific activity tracking that distinguishes pushes from steps.} that are currently available only through specialized wearables. To illustrate this feasibility, Figure~\ref{fig:propulsion_recog} shows synchronized accelerometer signals collected at 100Hz from an Apple Watch worn on the left wrist and from phones (iPhone 14 Pro) carried simultaneously in a trouser pocket and under the lap during repeated wheelchair propulsion by the first author. Propulsion cycles are clearly visible in the watch signal, with closely aligned peaks and troughs appearing in both phone signals. This alignment indicates that body-coupled phone placements capture propulsion-related dynamics similarly to wrist-worn devices, demonstrating the feasibility of propulsion tracking using phones alone without requiring additional wearables. Prior work has also shown that body-worn and chair-mounted IMUs can estimate the pose and movement patterns of wheelchair users~\cite{li2024wheelposer, barbareschi2018use}. Smartphones in body-coupled positions, paired with smartwatches and earbuds, can deliver comparable capabilities for everyday motion tracking and injury-prevention support~\cite{li2023breaking, mercer2006shoulder, fari2023shoulder, li2025optimizing}. 
Here, we note that because this example is based on synchronized signals from the first author, it should be interpreted as an initial feasibility demonstration. Future work should evaluate this direction with a broader sample of wheelchair users, a wider range of body-coupled placements, and subject-independent protocols, while comparing phone-based sensing against wrist-worn wearables or motion-capture references. Such an evaluation would help determine when phone-only sensing is sufficient and when multimodal combinations provide added value.

\begin{figure}[h]
\centering
\includegraphics[width=0.6\linewidth]{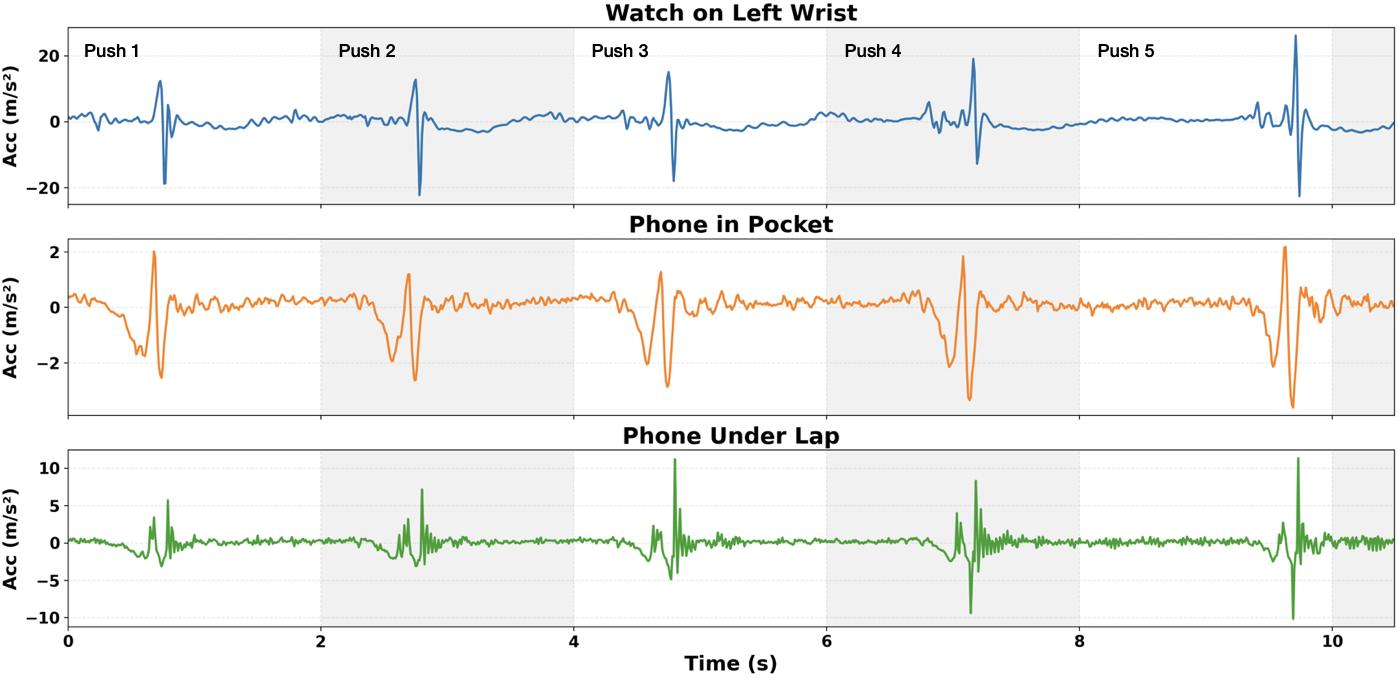}
\caption{Synchronized accelerometer signals from a watch (wrist) and phones (pocket, under the lap) during five manual pushes. The clear temporal alignment of peaks across devices demonstrates that body-coupled phones capture propulsion dynamics comparable to wrist-worn wearables.}
\Description{Three stacked line graphs showing accelerometer data over a 10-second period. The top graph shows data from a Watch on the Left Wrist, displaying five distinct spikes labeled Cycle 1 through Cycle 5. The middle graph shows data from a Phone in a Pocket. The bottom graph shows data from a Phone Under the Lap. All three graphs show spikes occurring at the exact same timestamps, indicating that the phone sensors are detecting the same propulsion movements as the wrist watch.}
\label{fig:propulsion_recog}
\end{figure}

\item \textbf{Environmental accessibility assessment from in-situ use.} When smartphones are coupled to the wheelchair, their motion sensors and microphones can further capture vibration and acoustic signatures that reflect travel surfaces and environmental conditions encountered during daily use. Figure~\ref{fig:terrain_sensing} illustrates spectrograms of vibration data captured by a phone's motion sensors at 100Hz from a frame-mounted holder as the first author traversed different surfaces in a manual wheelchair. The spectrograms reveal clear differences across terrain types: polished concrete exhibits low energy concentrated in lower frequencies, while tile and outdoor rough asphalt show progressively higher energy in high-frequency domains. Surface features such as connections and tactile paving are also visible as distinct patterns in the spectrograms.
Similarly, when the phone is forward-facing in a holder or resting in a cup holder, its camera can opportunistically capture brief, opt-in snapshots (or depth samples, where available) to document surface quality and accessibility barriers, complementing the inertial and audio cues.
Aggregated across users and trips, these signals could enable passive and scalable assessments of environmental accessibility from real-world usage data, reducing reliance on manual effort~\cite{li2024never, saha2019project}. Future work should evaluate this direction with data from more wheelchair users, a broader range of surfaces and barriers, and controlled studies that account for realistic variation in speed, wheelchair type, and mounting location. In-the-wild deployments would also be important for determining whether models developed in controlled settings remain reliable during everyday travel.

    \begin{figure}[h]
    \centering
    \includegraphics[width=0.8\linewidth]{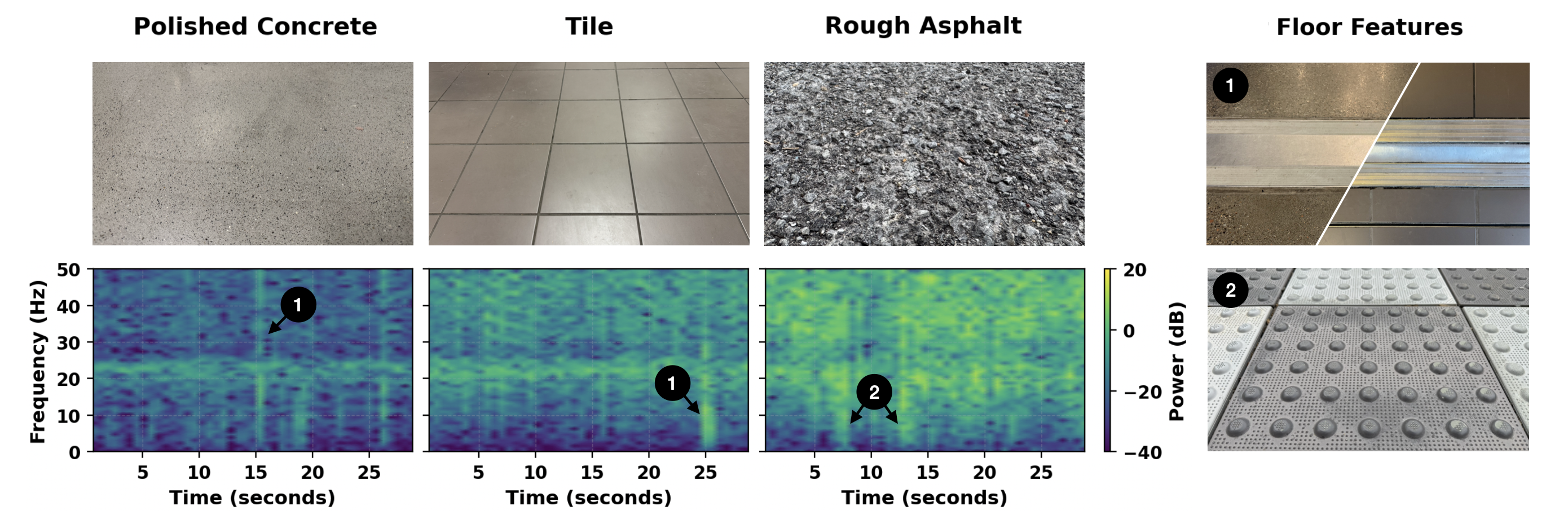}
    \caption{Vibration signatures of three distinct road surfaces. Spectrograms (0-50Hz) captured from a frame-mounted phone show increasing spectral energy as terrain shifts from polished concrete to rough asphalt. Transient environmental features are uniquely identifiable: (1) a metal floor transition appears as a distinct impulse, and (2) tactile paving creates a sustained high-energy pattern.}
    \Description{A composite figure comparing visual surfaces with their vibration data. The top row shows photos of Polished Concrete, Tile, and Rough Asphalt. Below each is a frequency spectrogram. Concrete shows dark blue (low energy). Tile shows some green. Asphalt shows bright yellow/green (high energy) across higher frequencies. On the far right, photos show specific floor features: a metal transition strip (labeled 1) and tactile paving bumps (labeled 2). Arrows on the spectrograms indicate where these specific features appear as distinct patterns in the data.}
    \label{fig:terrain_sensing}
    \end{figure}   

\end{enumerate}

\subsection{Ethical Considerations for Placement-Aware Systems}
While placement-aware and opportunistic sensing creates promising opportunities for inclusive mobile interaction, these approaches also raise ethical, privacy, and safety considerations. Because phones may be carried on the body, attached to wheelchairs, or positioned to capture aspects of the surrounding environment, sensing modalities such as motion, audio, and camera can reveal sensitive information about users' activities, routines, and mobility patterns, echoing tensions documented around sharing everyday activity data~\cite{li2023privacy}.
Designers should therefore treat consent, transparency, and user control as core requirements. Systems should clearly communicate what is being sensed, when sensing is active, and how data is processed and retained. In general, designs should favor data minimization, on-device processing where feasible, and user control over whether higher-risk sensing modalities are enabled.
Safety must also be considered alongside privacy. Context-aware interventions should not create new burdens or distract users during navigation. Ethical placement-aware design therefore requires balancing utility with privacy, autonomy, and physical safety, and future work should evaluate these tradeoffs directly with wheelchair users in real-world contexts.

\section{Limitations}
In this work, we explored the phone-carrying location preferences of wheelchair users through surveys and interviews. While our methods effectively identify what placements exist and why users choose them, they do not capture fine-grained quantitative behavioral data such as placement frequency, duration, or transition patterns throughout the day. Future work employing longitudinal methods such as diary studies or passive data logging over time, as demonstrated in prior research on the general population~\cite{dey2011getting, patel2006farther}, would complement our foundational understanding by modeling these temporal behavioral patterns.
Our study is also limited by its sample size and demographic diversity. All survey respondents were based in the United States and recruited through English-language social media communities and advocacy organizations, which may overrepresent individuals who are more digitally engaged and limit generalizability to other geographic and cultural contexts.
A larger and more heterogeneous participant pool may reveal new trends and variations in preferences. Moreover, all of our interview participants were full-time wheelchair users who could operate their phones independently. As such, our findings may not generalize to individuals with more severe motor impairments or those who use wheelchairs only part-time. Exploring the experiences of these groups would offer a more comprehensive understanding of diverse needs.
Further, this work did not account for wheelchair use in low-resource or informal settings, where our findings may have limited applicability~\cite{barbareschi2020bridging,vergunst2015you}.
Finally, our empirical demonstrations focused on manual wheelchairs and iOS devices due to equipment and testing constraints. While we expect the sensing feasibility and application domains to extend to power wheelchairs and other device ecosystems, specific signal patterns may differ. Future work should validate these opportunities across a wider range of wheelchair and mobile device types.

\section{Conclusion}

In this work, we present the first systematic characterization of phone-carrying practices among wheelchair users, revealing fundamental differences from general population behaviors with critical implications for mobile accessibility and context-aware computing. Through surveys with 91 and interviews with 15 wheelchair users, we documented a diverse ecosystem of carrying strategies that extend far beyond conventional pockets and bags to include wheelchair-specific solutions and adaptive body placements. We also explored a set of intrinsic and extrinsic factors that shape these behaviors and demonstrated how these carrying practices encode rich contextual information about users' situational needs, interaction intent, and mobility state, offering an untapped resource for mobile context-awareness. Taken together, these contributions advance mobile accessibility research by reframing phone carrying from a mundane logistical task to a lens for understanding the embodied experience of mobile technology use in wheelchair contexts. Through our design implications and future directions, we lay essential groundwork for developing inclusive tracking and adaptive applications that work with, rather than despite, the diverse ways wheelchair users integrate mobile devices into their daily lives, ultimately supporting more equitable access to the benefits of context-awareness.

\begin{acks}
The authors would like to thank all participants for their time and effort in this study. We extend our gratitude to Tong Wang for her assistance in refining the figures. Finally, we thank Yuyu Lin, all members of the AXLE Lab at CMU, and the anonymous reviewers for their valuable suggestions and feedback.
\end{acks}

\bibliographystyle{ACM-Reference-Format}
\bibliography{references}

\newpage
\appendix
\section{Supplementary Survey Results by Gender and Age}
\label{survey_age_gender}
\begin{table}[h]
\small
\renewcommand{\arraystretch}{1.1}
\begin{tabular}{lccc}
\hline
\textbf{Storage Location}                & \textbf{Man (42)}       & \textbf{Woman (46)}     & \textbf{Other (3)}    \\ \hline
Bag or Purse Attached to Wheelchair     & 35.7\% (15)        & 39.1\% (18)        & -                 \\
Between or On the Lap                   & 19.0\% (8)         & 32.6\% (15)        & 66.7\% (2)        \\
Phone Holder Mounted on Wheelchair      & 26.2\% (11)        & 19.6\% (9)         & 33.3\% (1)        \\
Trouser Pocket                          & 28.6\% (12)        & 13.0\% (6)         & 66.7\% (2)        \\
Crossbody Purse or Bag                  & 16.7\% (7)         & 13.0\% (6)         & -                 \\
Upper-Body Pocket                      & 9.5\% (4)          & 13.0\% (6)         & 33.3\% (1)        \\
Under the Lap                   & 14.3\% (6)         & 4.3\% (2)          & -                 \\
Pocket on Armband or Strap           & 7.1\% (3)          & 8.7\% (4)          & -                 \\
Cup Holder                              & 4.8\% (2)          & 8.7\% (4)          & -                 \\
Others                                  & 16.7\% (7)         & 8.7\% (4)          & 33.3\% (1)        \\ \hline
\multicolumn{4}{l}{Note: Participants could select up to three phone-carrying locations, so percentages do not sum to 100\%.} \\ \hline
\end{tabular}
\caption{Phone-carrying location preferences by gender.}
\label{tab:gender}
\end{table}

\begin{table}[h]
\small
\renewcommand{\arraystretch}{1.1}
\resizebox{\textwidth}{!}{%
\begin{tabular}{lccccc}
\hline
\textbf{Storage Location}               & \textbf{18–24 Years (12)} & \textbf{25–34 Years (31)} & \textbf{35–44 Years (29)} & \textbf{45–54 Years (12)} & \textbf{55+ Years (7)} \\ \hline
Bag or Purse Attached to Wheelchair     & 33.3\% (4)           & 38.7\% (12)          & 37.9\% (11)          & 25.0\% (3)           & 42.9\% (3)        \\
Between or On the Lap                   & 33.3\% (4)           & 35.5\% (11)          & 27.6\% (8)           & 16.7\% (2)           & -                 \\
Phone Holder Mounted on Wheelchair      & 33.3\% (4)           & 29.0\% (9)           & 20.7\% (6)           & 8.3\% (1)            & 14.3\% (1)        \\
Trouser Pocket                         & 25.0\% (3)           & 22.6\% (7)           & 13.8\% (4)           & 41.7\% (5)           & 14.3\% (1)        \\
Crossbody Purse or Bag                  & -                    & 16.1\% (5)           & 17.2\% (5)           & 8.3\% (1)            & 28.6\% (2)        \\
Upper-Body Pocket                      & 25.0\% (3)           & 9.7\% (3)            & 13.8\% (4)           & -                    & 14.3\% (1)        \\
Under the Lap                   & -                    & 12.9\% (4)           & 13.8\% (4)           & -                    & -                 \\
Pocket on Armband or Strap           & 8.3\% (1)            & 6.5\% (2)            & 3.4\% (1)            & 16.7\% (2)           & 14.3\% (1)        \\
Cup Holder                              & -                    & 3.2\% (1)            & 13.8\% (4)           & 8.3\% (1)            & -                 \\ 
Others                                  & 8.3\% (1)            & 19.4\% (6)           & 6.9\% (2)            & 16.7\% (2)           & 14.3\% (1)        \\\hline
\multicolumn{6}{l}{Note: Participants could select up to three phone-carrying locations, so percentages do not sum to 100\%.} \\ \hline
\end{tabular}
}
\caption{Phone-carrying location preferences by age group.}
\label{tab:age}
\end{table}

\begin{table}[h]
\small
\renewcommand{\arraystretch}{1.1}
\resizebox{\textwidth}{!}{%
\begin{tabular}{lccccc}
\hline
\textbf{Group} & \textbf{Bag or Purse} & \textbf{Between/On Lap} & \textbf{Phone Holder} & \textbf{Trouser Pocket} & \textbf{Crossbody Purse} \\ \hline
Man (42) &
35.7\% [23.0, 50.8] &
19.0\% [10.0, 33.3] &
26.2\% [15.3, 41.1] &
28.6\% [17.2, 43.6] &
16.7\% [8.3, 30.6] \\

Woman (46) &
39.1\% [26.4, 53.5] &
32.6\% [20.9, 47.0] &
19.6\% [10.7, 33.2] &
13.0\% [6.1, 25.7] &
13.0\% [6.1, 25.7] \\

Other (3) &
0.0\% [0.0, 56.2] &
66.7\% [20.8, 93.9] &
33.3\% [6.1, 79.2] &
66.7\% [20.8, 93.9] &
0.0\% [0.0, 56.2] \\ \hline
\multicolumn{6}{p{\linewidth}}{\footnotesize Note: Shown are the five most frequently selected phone-carrying locations overall, with 95\% Wilson confidence intervals. Wide intervals for smaller subgroups indicate limited precision in these descriptive estimates.} \\
\hline
\end{tabular}
}
\caption{Selected phone-carrying proportions by gender, with 95\% Wilson confidence intervals.}
\label{tab:gender_ci_key}
\end{table}

\begin{table}[h]
\small
\renewcommand{\arraystretch}{1.1}
\resizebox{\textwidth}{!}{%
\begin{tabular}{lccccc}
\hline
\textbf{Group} & \textbf{Bag or Purse} & \textbf{Between/On Lap} & \textbf{Phone Holder} & \textbf{Trouser Pocket} & \textbf{Crossbody Purse} \\ \hline
18--24 (12) &
33.3\% [13.8, 60.9] &
33.3\% [13.8, 60.9] &
33.3\% [13.8, 60.9] &
25.0\% [8.9, 53.2] &
0.0\% [0.0, 24.3] \\

25--34 (31) &
38.7\% [23.7, 56.2] &
35.5\% [21.1, 53.1] &
29.0\% [16.1, 46.6] &
22.6\% [11.4, 39.8] &
16.1\% [7.1, 32.6] \\

35--44 (29) &
37.9\% [22.7, 56.0] &
27.6\% [14.7, 45.7] &
20.7\% [9.8, 38.4] &
13.8\% [5.5, 30.6] &
17.2\% [7.6, 34.5] \\

45--54 (12) &
25.0\% [8.9, 53.2] &
16.7\% [4.7, 44.8] &
8.3\% [1.5, 35.4] &
41.7\% [19.3, 68.0] &
8.3\% [1.5, 35.4] \\

55+ (7) &
42.9\% [15.8, 75.0] &
0.0\% [0.0, 35.4] &
14.3\% [2.6, 51.3] &
14.3\% [2.6, 51.3] &
28.6\% [8.2, 64.1] \\ \hline
\multicolumn{6}{p{\linewidth}}{\footnotesize Note: Shown are the five most frequently selected phone-carrying locations overall, with 95\% Wilson confidence intervals. Wide intervals for smaller subgroups indicate limited precision in these descriptive estimates.} \\
\hline
\end{tabular}
}
\caption{Selected phone-carrying proportions by age group, with 95\% Wilson confidence intervals.}
\label{tab:age_ci_key}
\end{table}

\end{document}